\documentclass[aps,prl,twocolumn,superscriptaddress,nofootinbib]{revtex4-1}
\pdfoutput=1
\usepackage{graphicx}
\usepackage{epstopdf}
\usepackage{gensymb}
\usepackage{amsmath}
\usepackage{units}
\usepackage{hyperref}
\usepackage{upgreek}

\begin{document}

\title{Antenna-coupled photon emission from hexagonal boron nitride tunnel junctions }

\author{M. Parzefall}
\author{P. Bharadwaj}
\author{A. Jain}
\affiliation{Photonics Laboratory, ETH Z\"urich, 8093 Z\"urich, Switzerland}

\author{T. Taniguchi}
\author{K. Watanabe}
\affiliation{National Institute for Material Science, 1-1 Namiki, Tsukuba, 305-0044 Japan}

\author{L. Novotny}
\affiliation{Photonics Laboratory, ETH Z\"urich, 8093 Z\"urich, Switzerland}

\date{\today}

\begin{abstract}
The ultrafast conversion of electrical signals to optical signals at the nanoscale is of fundamental interest for data processing, telecommunication and optical interconnects. However, the modulation bandwidths of semiconductor light-emitting diodes are limited by the spontaneous recombination rate of electron-hole pairs, and the footprint of electrically driven ultrafast lasers is too large for practical on-chip integration. A metal-insulator-metal tunnel junction approaches the ultimate size limit of electronic devices and its operating speed is fundamentally limited only by the tunneling time. Here, we study the conversion of electrons -- localized in vertical gold-hexagonal boron nitride-gold tunnel junctions -- to free-space photons, mediated by resonant slot antennas. Optical antennas efficiently bridge the size mismatch between nanoscale volumes and far-field radiation and strongly enhance the electron-photon conversion efficiency. We achieve polarized, directional and resonantly enhanced light emission from inelastic electron tunneling and establish a novel platform for studying the interaction of electrons with strongly localized electromagnetic fields. \\

\end{abstract}


\maketitle

Optical nanoantennas are elements that couple free space radiation to material structures with length scales that are orders of magnitude smaller than the wavelength of a visible photon (400 to \unit[700]{nm})~\cite{bharadwaj09a,alu08a,greffet10a,novotny11a,biagioni12a}. 
Incoming radiation is converted into localized surface plasmon polaritons (LSPPs) -- time-harmonic oscillations of the free electron gas 
-- that concentrate electromagnetic energy into ultrasmall volumes \cite{schuller10}. Many optical antennas owe their design to their macroscopic radiofrequency predecessors like linear dipole antennas \cite{muehlschlegel05,alu08a}, Yagi-Uda antennas \cite{curto10}
or bow-tie antennas \cite{kinkhabwala09}. \\[-1.5ex] 

Classical antennas are used to generate radiofrequency (RF) waves by driving them at the respective frequency electrically or -- in reverse -- to generate RF electrical signals from incoming electromagnetic radiation. Optical antennas on the other hand have been mostly operated on a ``light-in / light-out'' basis \cite{novotny11a}. This paradigm has recently started to shift towards the integration of optical antennas in optoelectronic devices for photovoltaics \cite{atwater10}, photon detection \cite{tang08,knight11}  and
surface plasmon excitation \cite{huang14}. Here we report on the realization of ultrafast solid-state light-emitting tunnel devices based on arrays of electrically driven optical antennas. \\[-1.5ex] 

\begin{figure}
	\includegraphics{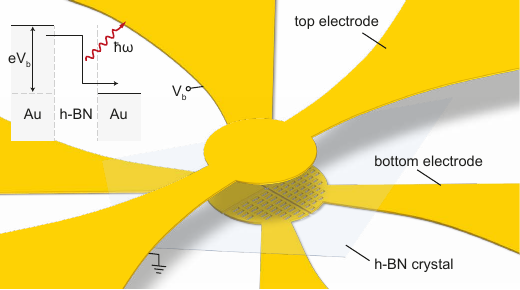}
	\caption{\label{fig1}Expanded illustration of the sample configuration. Devices consist of a vertical stack of segmented, nanostructured gold bottom electrodes, few-layer hexagonal boron nitride (h-BN) and a common gold top electrode. We observe and analyze light emission due to inelastic electron tunneling (see inset), controlled by the optical, geometrically defined, properties of the devices. The inset shows the generation of a photon with energy $\hbar\omega$ by inelastic electron tunneling.}
\end{figure}

\begin{figure*}
	\centering
	\includegraphics{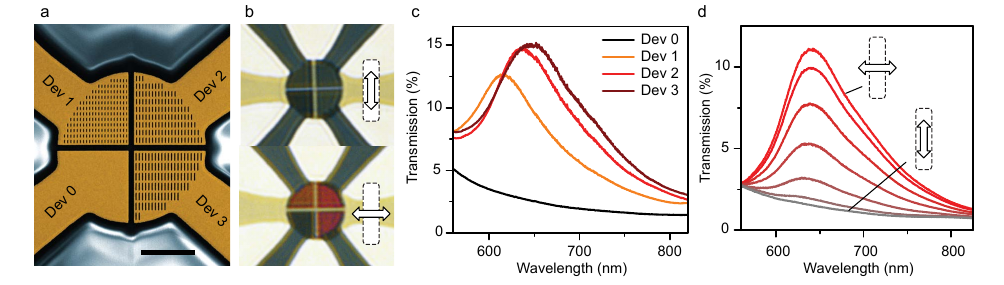}
	\caption{\label{fig2} Structural and optical properties of the tunnel devices. (a) False color SEM image of a nanostructured bottom electrode. Sizes of the slots are $\unit[150 \times 50]{nm^2}$, $\unit[250 \times 50]{nm^2}$ and $\unit[350 \times 50]{nm^2}$ for Devices 1, 2 and 3, respectively. Scale bar $\unit[2.5]{\upmu m}$. (b) Polarized optical transmission micrographs of the final device. Light polarized parallel to the short axis of the slots generates distinct color variations amongst the different electrodes. The colors disappear for light polarized along the long axis. Transmission through the unstructured reference Device 0 does not depend on polarization. (c) Transmission spectra of the four devices, quantifying the qualitative observation from (b). The antenna arrays give rise to characteristic spectral resonances. (d) Transmission spectra  of Device 3 as a function of polarizer angle from $\unit[0]{\degree}$ (parallel to the short axis) to $\unit[90]{\degree}$ (parallel to the long axis) in steps of $\unit[15]{\degree}$. }
\end{figure*}

Applying a voltage to a metal-insulator-metal junction yields a current due to electrons tunneling through the insulator from occupied states in one electrode to unoccupied states in the other electrode. 
The majority of electrons tunnel elastically, maintaining their energy during the tunneling process. The excess energy of the ``hot electron'' is subsequently thermalized. In 1976, Lambe and McCarthy found that surface plasmon modes in metal-insulator-metal (MIM) tunnel junctions increase the 
probability of inelastic tunneling, a process in which a tunneling electron excites a surface plasmon, c.f. inset of Fig. \ref{fig1}, which may subsequently decay into far-field radiation \cite{lambe76}. This effect has since been studied experimentally in macroscopic solid-state tunnel junctions \cite{kirtley81,sparks92}, 
and the scanning tunneling microscope (STM)
 \cite{gimzewski88,schull09,chen09b,bharadwaj11b,zhang13},
  as well as theoretically \cite{davis77,rendell81,johansson90,persson92,aizpurua00}. One major appeal of investigating inelastic electron tunneling is its potential speed. Since it does not rely on intermediate excitations such as electron-hole-pairs, the response time of such devices is fundamentally only limited by the tunneling time of electrons through the junction, a process which takes place on a femtosecond timescale \cite{landauer94,shafir12}. This enables tunnel devices to operate at optical frequencies \cite{ward10}.  \\[-1.5ex] 

The tunnel barrier of top-down fabricated devices is commonly based on the oxides grown on one of the electrodes (e.g. aluminum oxide) or on insulating films deposited by atomic-layer deposition (ALD) or sputtering. Unfortunately, due to grain formation and defects, these tunnel barriers are unstable and operation over extended time at ambient conditions is not possible \cite{kirtley81}. Here, we investigate light generation via inelastic electron tunneling in microscopic tunnel devices comprised of gold (Au) and hexagonal boron nitride (h-BN). h-BN provides a stable, high-quality tunnel barrier due to its crystallinity \cite{lee11b,britnell12a} and a large bandgap of $\sim \, \! \unit[6]{eV}$ \cite{watanabe04} (see Supplementary Information, Section 5.1). We find that light emission can be enhanced and spectrally controlled by nanostructuring one of the electrodes. Light emission and optical properties of the device are found to be closely related. Furthermore, we demonstrate modulation of the emitted light at frequencies up to \unit[1]{GHz}.  \\[-1.5ex] 

Fig. \ref{fig1} illustrates the sample configuration. We photolithographically define a circular gold bottom electrode (\unit[1]{nm} Ti / \unit[50]{nm} Au) that is subsequently nanostructured by focused ion beam milling (FIB). The bottom electrode is segmented into four quarter-circle electrodes, which are wired individually as shown in Fig. \ref{fig2}a. One electrode remains unstructured (Device 0) to serve as a reference, while the remaining three electrodes are structured into arrays of optical antennas in the form of rectangular slots. The distance between slots is kept constant at $\sim \, \!  \unit[100]{nm}$ along both the long and short side of the rectangles while the nominal size of the slots is $\unit[150 \times 50]{nm^2}$, $\unit[250 \times 50]{nm^2}$ and $\unit[350 \times 50]{nm^2}$ for Devices 1, 2 and 3, respectively. Exfoliated few-layer h-BN (here: 9 atomic layers, or $\sim \, \! \unit[3]{nm}$) is transferred on top of the bottom electrode. The device is finalized by the fabrication of a common gold top electrode (\unit[15]{nm} Au) via photolithography. For details on the fabrication process, see Methods and Supplementary Information, Section 1.  \\[-1.5ex] 

The optical properties of the final devices are shown in Fig. \ref{fig2}b-d. The transmission of light through the devices strongly depends on its (linear) polarization, as shown in Fig. \ref{fig2}b. Observing light polarized along the short axis of the slots results in a strong color contrast between devices. This contrast vanishes when observing light polarized along the long axis. 
The colors of the electrodes suggest that the slots enhance transmission in a certain spectral range. This observation is quantified in the transmission spectra $T (\lambda)$ shown in Fig. \ref{fig2}c. 
All structured devices exhibit characteristic resonances and enhanced transmission at wavelengths corresponding to the observed colors in Fig. \ref{fig2}b. The transmission peak shifts towards longer wavelengths as the aspect ratio of the slots is increased in accordance with Refs. \cite{kleinkoerkamp04,garciavidal05}. Fig. \ref{fig2}d shows the change in the transmitted spectrum as a function of polarizer angle. The resonance disappears when the analyzer is oriented along the long axis of the slots. The transmission follows a $\cos ^2 \theta$ dependence (see Supplementary Information, Section 2) \cite{gordon04}.  \\[-1.5ex] 

\begin{figure*}
	\includegraphics{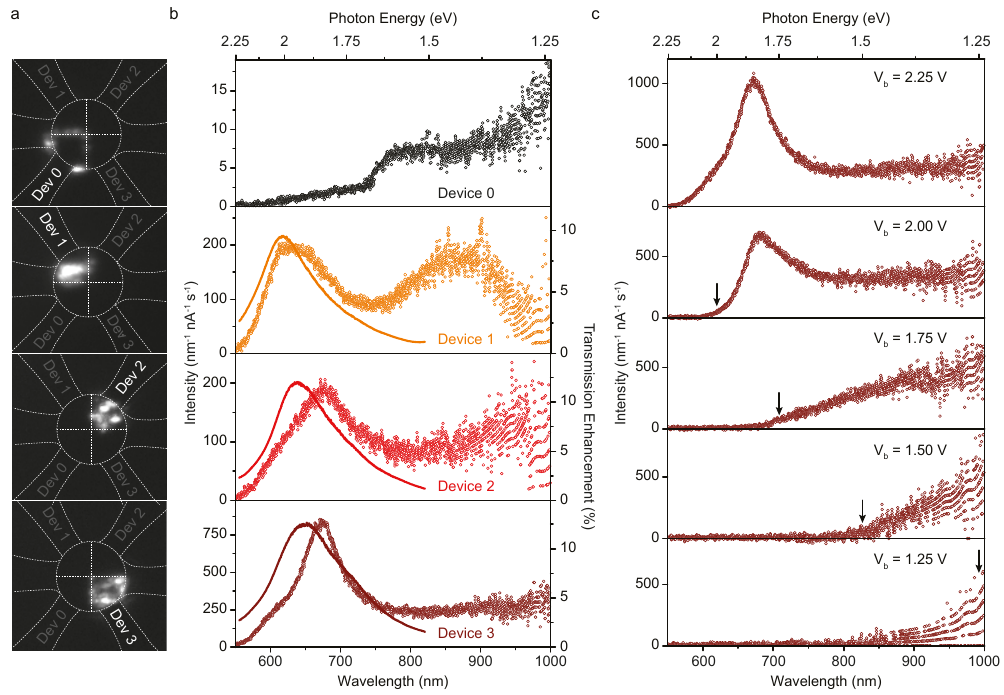}
	\caption{\label{fig3}Light emission from antenna-coupled tunneling devices. (a) EMCCD images of the light emitted from the four devices at a constant applied bias of $V_{\rm b} = \unit[2.5]{V}$. (b) Spectral intensity of the emitted light for all four devices at $V_{\rm b} = \unit[2.5]{V}$ (open circles) in comparison with the optical transmission enhancement, i.e. $\Delta T_i ( \lambda ) = T_i (\lambda) - T_0 (\lambda)$ (solid lines). Resonance peaks are observed in all structured devices. The peaks shift towards longer wavelenghts with increasing aspect ratio of the slots. (c) Light emission spectra for Device 3 as a function of the applied bias. The arrows indicate the respective cut-off photon energy given by $\hbar \omega = | e V_{\rm b} |$.}
\end{figure*}

Next, we discuss the properties of photon emission induced by inelastic electron tunneling. The top-most image in Fig. \ref{fig3}a shows a real-space map of the light emitted from the unstructured reference device, recorded with a constant tunnel bias voltage of $V_{\rm b} = \unit[2.5]{V}$. 
Light emission is localized to the edges of the device. The spectrum of the emitted light, shown in Fig. \ref{fig3}b, is broad and without distinct features, falling off continuously towards shorter wavelengths. In the absence of slots, as is the case for the reference device, inelastically tunneling electrons predominantly interact with surface plasmon polaritons (SPPs) associated with the metal-insulator-metal (MIM) configuration \cite{miyazaki06,chen13}. 
In the absence of discontinuities, these modes cannot couple to free-space radiation because of the large momentum mismatch (c.f. Supplementary Information, Section 3.2). However, scattering due to surface roughness or edges provides the momentum required to overcome this mismatch. 
\\[-1.5ex] 

\begin{figure*}
	\includegraphics{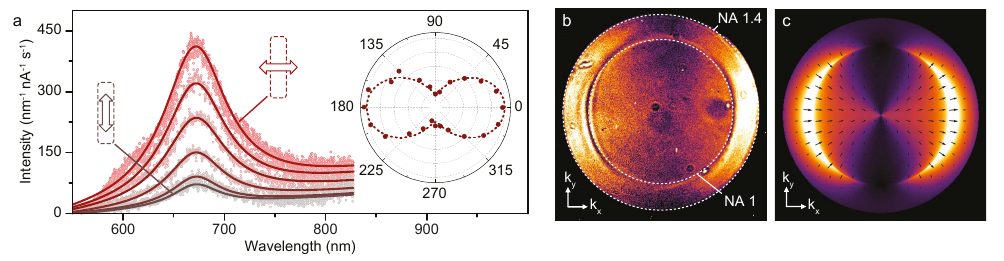}
	\caption{\label{fig4}Dipolar characteristics of the emitted light. (a) Light emission spectra of Device 3 at $V_{\rm b} = \unit[2.5]{V}$ as a function of the analyzing polarizer angle $\theta$ from $\unit[0]{\degree}$ (parallel to the short axis) to $\unit[90]{\degree}$ (parallel to the long axis) in steps of $\unit[15]{\degree}$. The superimposed curves are Lorentzian fits on a linear background. The inset shows a polar plot of the spectrally integrated light emission intensity fitted by a $\cos ^2 \theta$ function. (b) Photon momentum distribution in the backfocal plane of the objective for Device 3 at $V_{\rm b} = \unit[2.5]{V}$. (c) Analytically calculated  photon momentum distribution for a magnetic dipole $m_y$ in air oriented parallel to and at a distance of \unit[25]{nm} from a glass ($n=1.52$) interface (objective NA 1.4). Vectors indicate direction and magnitude of the real part of the electric field at the respective vector positions. The dark ring at an angle slightly larger than NA 1.0 observed experimentally is not reproduced by analytical calculations. This is most likely caused by coupling to surface plasmon polaritons at the gold-air interface, which is not accounted for in the calculation.}
\end{figure*}

In contrast, all devices with nanostructured bottom electrodes show light emission from the entire device area (Fig. \ref{fig3}a). As seen in the corresponding spectra in Fig. \ref{fig3}b, the antenna-coupled devices exhibit strongly enhanced light emission intensity, especially in the spectral domain of the transmission peaks discussed previously.  The external electron-to-photon conversion efficiency is increased by two orders of magnitude from $QE_{\mathrm{D0}} \approx \, \! 4 \cdot 10^{-7}$ to $QE_{\mathrm{D3}} \approx \, \! 2.5 \cdot 10^{-5}$ at $V_b = \unit[2.5]{V}$. A direct comparison of the transmission and emission spectra shows good agreement in spectral width and shape. However, the peaks observed in the light emission are red-shifted compared to the peaks in transmission. Numerical simulations, which will be discussed later on, suggest that this spectral shift is caused by the spectral dependence of electron-radiation coupling mediated by the slot antennas. The spatial non-uniformity of the emitted light intensity is due to fabrication-related contamination at the Au--h-BN interfaces and roughness of the polycrystalline electrodes to which the atomically flat h-BN crystal cannot conform perfectly \cite{britnell12a}. Photon emission can be made spatially more uniform by using monocrystalline or template-stripped gold electrodes \cite{nagpal09,huang10b}.  \\[-1.5ex] 

As reported by Lambe and McCarthy, inelastic electron tunneling results in broad-band light emission with a high-frequency cutoff given by
\begin{equation}
\hbar \omega_{\mathrm{max}}=\left| e V_{\rm b}\right|,
\label{eq1}
\end{equation}
with $\hbar \omega_{\mathrm{max}}$ being the photon energy, $e$ the electron charge and $V_b$ the tunnel bias voltage \cite{lambe76}. According to equation \eqref{eq1}, the antenna-specific resonances should only be excitable if the applied bias exceeds the energy of the antenna mode. As shown in Fig. \ref{fig3}c, this behaviour is indeed observed. The figure renders light emission spectra for Device 3 as a function of applied bias $V_{\rm b}$. The experimental data shows good agreement with equation \eqref{eq1}.  \\[-1.5ex] 

To understand the role of the slot antennas in the light generation process, we studied  the polarization and radiation characteristics of the emitted light. Fig. \ref{fig4}a shows emission spectra for Device 3 as a function of polarizer angle. The light intensity is maximized when the polarizer is oriented parallel to the short axis of the slots. 
For a polarizer angle parallel to the long axis of the slots we observe a $>$ \unit[80]{\%} extinction. 
Fig. \ref{fig4}b shows the photon momentum distribution in the back focal plane of the microscope objective (NA 1.4). Light is preferentially emitted in $x$-direction, i.e. parallel to the short axis of the slots. The measured polarization and the back focal plane intensity distribution qualitatively agree with the radiation generated by a magnetic dipole over a dielectric substrate~\cite{novotny12} and oriented along the long axis of the slots (c.f. Fig. \ref{fig4}c). These results are in agreement with the notion that slots can be viewed -- in terms of Babinet's principle --  as the complement of linear rod antennas, radiation from which is dominated by an electric dipole mode along the long axis of the rod \cite{yang14}. However, as will be discussed later, slots also support electric dipole resonances parallel to their short axis.  \\[-1.5ex] 

\begin{figure}
	\includegraphics{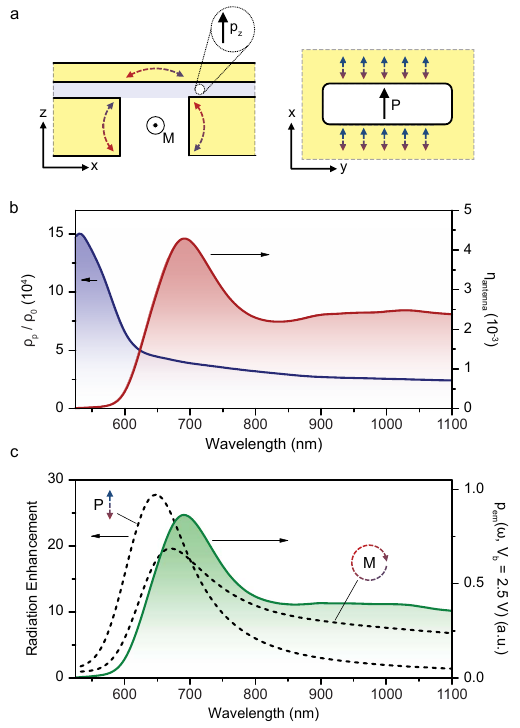}
	\caption{\label{fig5}  Finite-element calculations and device emission spectrum. (a) Graphical illustration of dipole placement ($p_z$) and  current distributions (dashed arrows) associated with magnetic (${\rm M}$, left) and electric (${\rm P}$, right) dipole resonances. (b) Normalized local density of electromagnetic states $\rho_{\rm p} / \rho_0$ and corresponding radiation efficiency $\eta_{\rm antenna}$ calculated for a $p_z$ dipole located in the center of the MIM gap and at a distance of \unit[5]{nm} from the antenna. (c) Mode spectra of ${\rm P}$ and ${\rm M}$ obtained by calculating the radiation enhancement $P_{\rm r} / P_{\rm 0}$ of elementary dipoles $p_x$/$m_y$, located as indicated in (a). In direct comparison we show the spectrum of emitted power $p_{\rm em} (\omega, V_{\rm b})$ for a bias of $V_{\rm b} = \unit[2.5]{V}$. The size of the slot is $\unit[150 \times 50 \times 50]{nm^3}$.}
\end{figure}

To establish an 
understanding of the electron-to-photon conversion efficiency and its spectral dependence, we express the spectrum of the emitted power 
as
\begin{equation}
p_{\rm em}(\omega, V_{\rm b}) \;=\; \hbar\omega \;\Gamma_{\rm e-p}(\omega, V_{\rm b}) \; \eta_{\rm antenna}(\omega)
\label{eq2}
\end{equation}
with $\Gamma_{\rm e-p}$ being the electron-to-plasmon conversion rate and  $\eta_{\rm antenna}$ the radiation efficiency of the antenna. $\Gamma_{\rm e-p}$ is described by Fermi's ``golden rule'' \cite{persson92,chen09b,schneider13} and depends on the applied bias $V_{\rm b}$, the electronic density of states in the electrodes as well as the number of plasmonic modes that the electrons can couple to~\cite{aizpurua00}. The latter corresponds to the local density of electromagnetic states (LDOS) in the MIM tunnel gap. The radiation efficiency $\eta_{\rm antenna}$ denotes the efficacy of the slot antennas at converting the energy of MIM plasmons into free space radiation. We next discuss the spectral dependence of $\Gamma_{\rm e-p}$ and $\eta_{\rm antenna}$ for a $\unit[150 \times 50 \times 50]{nm^3}$ slot antenna; see Supplementary Information, Section 4 for additional data. \\[-1.5ex] 

We derive the LDOS seen by the tunneling electrons by numerically calculating the total dissipated power  $P_{\rm tot}$ of a point dipole $p_z$ placed in the tunnel junction. 
The LDOS $\rho_{\rm p}$ is then obtained by~\cite{novotny12}
\begin{equation}
\rho_{\rm p} = \rho_0 \cdot \frac{P_{\rm tot}}{P_0} ,
\end{equation}
where $\rho_0 = \omega^2 \pi^{-2} c^{-3}$ is the vacuum LDOS and $P_0$ is the radiated power of a dipole of equal dipole moment in a homogeneous dielectric environment. 
Fig. \ref{fig5}b shows the normalized LDOS $\rho_{\rm p} / \rho_0$ for a dipole placed at a distance of $5\,$nm from the edge of an antenna slot (c.f. Fig.~\ref{fig5}a). A large fraction of the LDOS is associated with MIM-SPP modes, which are ultimately dissipated to heat in the absence of antenna coupling (see Supplementary Information, Section 4.3).
 \\[-1.5ex] 

In terms of the radiated power $P_{\rm r}$, the radiation efficiency can be expressed as
\begin{equation}
\eta_{\rm antenna} = \frac{P_{\rm r}}{P_{\rm tot}} \; .
\end{equation}
Fig. \ref{fig5}b renders $\eta_{\rm antenna}$ as a function of wavelength for the dipole near the edge of an antenna slot. We find that the antenna converts SPP modes  to propagating radiation most efficiently in a narrow spectral range centered at $\lambda=\unit[690]{nm}$. This antenna resonance can be associated with radiative magnetic and electric dipole modes, as discussed below. By varying the lateral position of $p_z$ in the tunnel gap we can map out the total LDOS $\rho_{\rm p}$ as well as the radiative LDOS $\rho_{\rm rad} = \rho_{\rm p} \cdot \eta_{\rm antenna}$ as a function of proximity to the antenna slot. 
The conversion of SPP modes to free-propagating radiation is most efficient for dipoles that are close to the antenna slot (see Supplementary Information, Section 4.3). Finally, this allows us to calculate the spectrum of the emitted power $p_{\rm em}(\omega, V_{\rm b})$. The modeled spectrum for an applied bias of $V_{\rm b} = \unit[2.5]{V}$ is shown in Fig. \ref{fig5}c. The spectral shape 
agrees well with our measurements shown 
in Fig.~\ref{fig3}b. For further details of the model and comparison to experimental data, see Supplementary Information, Section 4.5. \\[-1.5ex] 

\begin{figure*}
	\includegraphics{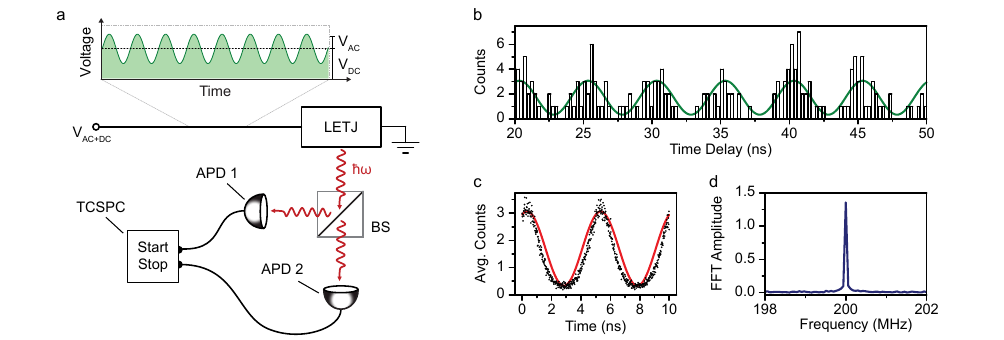}
	\caption{\label{fig6} Frequency modulation of light emission. (a) A light emitting tunnel junction (LETJ) is driven by a radio frequency (RF) signal of the form $V (t) = V_{\mathrm{DC}} + V_{\mathrm{AC}} \sin \left( 2 \pi f t \right)$. Emitted photons are directed to two avalanche photo diodes (APDs) via a beamsplitter (BS). A TCSPC module is used to create a histogram of the relative photon arrival times. (b) Excerpt of a TCSPC histogram of light emission modulated at $f = \unit[200]{MHz}$ ($V_{\mathrm{DC}} = \unit[1.5]{V}$, $V_{\mathrm{AC}} = \unit[0.5]{V}$) overlayed with a sine function of frequency $f$. (c) Histogram data from (b) folded back onto and averaged over two oscillation periods in comparison with a sine wave of frequency $f$. (d) Fast Fourier transform (FFT) spectrum of the histogram exhibiting a peak at the modulation frequency.}                 
\end{figure*}

In order to understand the experimentally observed spectral shift between transmission and light emission resonances, as well as the magnetic dipole character of light emission, it is necessary to identify the nature and symmetry of electromagnetic modes supported by slot antennas.
Radiation from an aperture can be described by a multipole series, with the leading terms for small apertures being magnetic and electric dipoles. 
As illustrated in Fig. \ref{fig5}a, we find that the dominating modes supported by slot antennas are an electric dipole mode along the short axis (${\rm P}$), and a magnetic dipole mode parallel to the long axis (${\rm M}$) of the slot, in agreement with recent cathodoluminescence measurements~\cite{coenen14}. 
Resonances of both magnetic and electric dipolar modes red-shift with increasing slot aspect ratio, in accordance with both experimental transmission and light emission results (see Supplementary Information, Section 4.1). The dashed curves in Fig. \ref{fig5}c show numerically obtained mode spectra for ${\rm P}$ and ${\rm M}$.  
The ${\rm M}$ resonance is red-shifted with respect to the ${\rm P}$ resonance. Furthermore, the ${\rm M}$ resonance is asymmetric with a long-wavelength tail due to strong coupling to MIM-SPPs, the damping of which increases with decreasing wavelength. \\[-1.5ex] 

Returning back to equation \eqref{eq2}, we now realize that the spectral dependence of the antenna-coupled tunneling device is dictated by the intricate interaction of tunneling electrons and electromagnetic modes supported by the slot antennas. 
As discussed earlier, our experimental results indicate that the enhanced light emission is mediated primarily by the magnetic dipolar antenna mode. This observation is substantiated by the spectral overlap of the emission spectrum $p_{\rm em}(\omega, V_{\rm b})$ with $\rm M$, as well as by the symmetry of this mode which favors field localization to the gap region. On the other hand, again for reasons of symmetry, linearly polarized far-field radiation primarily couples to in-plane electric dipole modes, i.e. the resonant enhancement of transmission through the device is mediated by the electric dipole mode ${\rm P}$. These conclusions are supported by the observed spectral shift between the transmission and emission peaks discussed earlier, which we also find between electric dipole mode spectrum ${\rm P}$ and emission spectrum (c.f. Fig. \ref{fig5}c).  \\[-1.5ex] 

In a final experiment, we demonstrate the feasibility of ultrafast temporal modulation of light emission from h-BN tunnel junctions. As depicted in Fig. \ref{fig6}a, we drive the device with an RF signal of frequency $f$, and record a histogram of the interphoton arrival times in the resulting photon stream by employing time-correlated single-photon counting (TCSPC). Fig. \ref{fig6}b shows an excerpt from such a histogram recorded for a modulation frequency of $f=\unit[200]{MHz}$, demonstrating that the photon stream is indeed time-modulated at the same frequency $f$. This can be seen unambiguously by averaging and Fourier-transforming the raw histogram, as shown in Figs. \ref{fig6}c,d. Following this approach, we have modulated photon emission from our devices at frequencies ranging from $f=\unit[10]{MHz}$ to $f=\unit[1]{GHz}$. For a detailed discussion of the method and further results, see Supplementary Information, Section 5.2. \\[-1.5ex]  

Further studies will be aimed at the increase of efficiencies through alternative antenna designs (e.g. patch antennas~\cite{bigourdan14} or nanocube antennas~\cite{akselrod14}) and improvements in fabrication and materials. For example,  MIM propagation lengths may be increased through the use of monocrystalline metals~\cite{huang10b} or alternative materials \cite{khurgin10,boltasseva11,tassin12}. Higher efficiencies and spectrally narrower resonances can also be achieved through MIM gap resonances.
Equivalent circuit models  will help to improve impedance matching between the tunnel junction and free-space, thereby improving the device efficiency further~\cite{alu08a,greffet10a}. \\[-1.5ex] 

In conclusion, we have demonstrated optical antenna mediated conversion of electronic into optical energy in a solid-state system, and its modulation at frequencies up to \unit[1]{GHz}. The vertical design of slot antennas allows for sub-nanometer control of gap-size and configuration, a goal which is difficult to achieve in traditional in-plane antenna designs. We find that arrays of slot antennas exhibit geometrically tunable, polarization-sensitive resonances that facilitate enhanced device transmission. While light emission from unstructured devices is weak and limited to the edges of the device, slot antenna arrays emit from the entire device area. The emitted light is strongly polarized along the short axis of the slots in correspondence with the polarization dependence of the transmission.  The exploration of junction configurations based on van der Waals heterostructures \cite{geim13} with tailored electronic properties may lead to significant improvements in efficiencies and enable novel device functionalities. By reciprocity, the geometry presented in this study is also capable of converting far-field radiation into localized energy, which is of interest for photovoltaics, photodetection and optical sensors.\\



\section{Methods}

\subsection{Sample Fabrication}

Devices are fabricated on commercially available glass coverslips ($\unit[22 \times 22 \times 0.13]{mm^3}$). Both bottom and top electrodes were fabricated by standard UV photolithography, electron-beam evaporation of titanium and gold targets, and subsequent lift-off. Bottom electrodes were nanostructured using a dual-beam FEI FIB Helios 600i. The top electrode of the device used in the frequency-modulation experiments was fabricated using a silicon nitride membrane shadow mask prepared by FIB milling. \\[-1.5ex]

h-BN crystals were grown as described in Ref. \cite{taniguchi07}. Few-layer h-BN atomic crystals were exfoliated on silicon wafers with a \unit[280]{nm} oxide layer using the scotch-tape micromechanical cleavage technique \cite{novoselov04a,novoselov05b}. Crystals were identified by means of optical microscopy as well as atomic force microscopy using a Bruker Innova AFM in tapping mode.  \\[-1.5ex]

The transfer was carried out as follows: 
The Si/SiO$_2$ substrate containing the h-BN crystals is partially covered with a commercial PDMS film (Gel-Film\textregistered\  PF-40-X4 supplied by Gel-Pak) in such a way that small margins along the edges of the substrate are left exposed.
The stack is subsequently floated on deionized water, the exposed (hydrophilic) Si/SiO$_2$ surface is wetted such that a meniscus is formed around the hydrophobic PDMS and heated to $\unit[90]{\degree C}$. After $\sim \, \! \unit[30]{min}$ the stack is completely immersed in water and the PDMS film is peeled-off under water. During this process the majority of h-BN crystals gets transferred from the Si/SiO$_2$ to the PDMS. The PDMS film, carrying the h-BN crystal, is subsequently aligned manually with the bottom electrode using a SUSS MJB4 mask aligner, attached to a quartz carrier, and brought into contact with the bottom electrode. As the PDMS film is released from the sample, the h-BN crystal gets transferred onto the bottom electrode. See Supplementary Information for AFM and optical microscope images of the transfer process.    \\[-1.5ex]

Samples were annealed in a tube furnace under a $\unit[400]{sccm}$ flow of $\unit[95]{\%}$ Argon and $\unit[5]{\%}$ Hydrogen at $\unit[300]{\degree C}$ for $\sim \unit[3]{h}$ before and after transfer. \\[-1.5ex]

\subsection{Sample Characterization}

All measurements were carried out on a customized, inverted Nikon TE300 microscope. For transmission measurements light from a supercontinuum source (NKT SuperK EXTREME) was focused onto the devices through a top objective (50x, NA 0.8). Transmitted light as well as light emitted by the devices was collected using an oil-immersion objective (100x, NA 1.4). Spectra of the transmitted as well as emitted light were acquired with an Acton SpectraPro 300i spectrometer. Transmission spectra are normalized by spectra acquired on glass, which is assumed to correspond to the theoretical value of \unit[96]{\%}. Units of light emission spectra are not arbitrary. They are normalized by the system efficiency, acquisition time and the average current flowing through the device during acquisition. The system response was calibrated using an OceanOptics HL-2000-CAL calibration lamp. Real space as well as back-focal plane images of the emitted light were acquired using an Andor iXon Ultra EMCCD camera. As a DC voltage source and for electrical measurements we used a Keithley 2602B source meter. The RF signal is generated by a Hewlett Packard 8648B signal generator. TCSPC histograms are acquired using a PicoQuant PicoHarp 300 module as well as two Excelitas Technologies SPCM-AQRH-14 APDs. \\[-1.5ex]

\subsection{Numerical Simulations}

Numerical finite-element simulations were carried out with COMSOL Multiphysics 4.4. We assumed refractive indices of 1.52 for glass and 1.8 for h-BN \cite{levinshtein01}. Empirical values were used for the dielectric function of gold \cite{johnson72}. The \unit[1]{nm} adhesion layer of the lower electrode was not taken into account. Full-field simulations of the electromagnetic fields were performed for electric / magnetic point dipoles at the positions mentioned in the main text. The radiated power was extracted by integrating the time-averaged Poynting vector over the simulation boundaries in the lower and upper half-spaces. The non-radiative energy loss rate is calculated by integrating the total power dissipation density within the metallic domains. Perfectly matched layers were used as simulation boundaries.


\section{Acknowledgments}
The authors thank Zachary J. Lapin for sample fabrication by focused ion beam milling, Karol Luszcz for support regarding the frequency modulation experiments as well as Mark Kasperczyk and Dieter W. Pohl for helpful discussions. Funding by the NCCR-QSIT program and the Swiss National Science Foundation (grant 200021\_149433) is greatly appreciated. We further acknowledge the use of facilities at the FIRST Center for Micro- and Nanotechnology as well as the Scientific Center for Optical and Electron Microscopy (ScopeM) at ETH Z\"urich. K.W. and T.T. acknowledge support from the Elemental Strategy Initiative conducted by the MEXT, Japan. T.T. acknowledges support a Grant-in-Aid for Scientific Research on Grant 262480621 and on Innovative Areas "Nano Informatics" (Grant 25106006) from JSPS.\\[1.5ex]

\section{Author Contributions}
L.N., P.B. and M.P. conceived the research. M.P. fabricated the samples and carried out the numerical simulations. M.P. and P.B. measured the samples. A.J. developed the
transfer technique. K.W. and T.T. synthesized the h-BN crystals. M.P, P.B. and L.N. discussed results and co-wrote the paper. \\[-1.5ex]

\end{document}